\documentclass[aps,prl,twocolumn,amsmath]{revtex4}
\usepackage{graphicx}
\usepackage{dcolumn}
\usepackage[mathcal]{euscript}
\usepackage{color}
\usepackage[dvipsnames]{xcolor}
\usepackage{hyperref}

\def\beq{\begin{equation}}
\def\eeq{\end{equation}}

\begin{document}


\title{Off-axis dipole forces in optical tweezers by an optical analog of the {Magnus} effect}

\author{Robert J.C. Spreeuw}
\affiliation{Van der Waals-Zeeman Institute, Institute of Physics, University of Amsterdam, \\
PO Box 94485, 1090 GL Amsterdam, The Netherlands}

\date{\today}

\begin{abstract}

It is shown that a circular dipole can deflect the focused laser beam that induces it, and will experience a corresponding transverse force. Quantitative expressions are derived for  Gaussian and angular tophat beams, while the effects vanish in the plane-wave limit. The phenomena are analogous to the Magnus effect pushing a spinning ball onto a curved trajectory. The  optical case originates in the coupling of spin and orbital angular momentum of the dipole and the light.
In optical tweezers the  force causes off-axis displacement of the trapping position of an atom by a spin-dependent amount up to $\lambda/2\pi$, set by the direction of a magnetic field. 
This suggests direct methods to demonstrate and explore these effects, for instance to induce spin-dependent motion.

\end{abstract}



\maketitle


A common practice in many branches of sports is  
to send a ball onto a curved trajectory by giving it a spin. In this famous example of the Magnus effect \cite{magnus_ueber_1853} the spinning ball deflects the stream of air around it, and is pushed sideways by the reaction force perpendicular to its forward velocity. 
In analogy, we may ask if a rotating dipole in an atom may  similarly deflect a beam of light, and thereby be pushed by a force perpendicular to the light beam. 
C.G.\ Darwin already remarked that for circular dipoles ``\dots the wave front of the emitted radiation faces not exactly away from the origin, but from a point about a wave-length away from it.''\cite{darwin_notes_1932}. 
A recent experiment confirmed that an atomic circular dipole can indeed appear to be displaced from its true location, due to the emitted spiral-shaped wavefront \cite{araneda_wavelength-scale_2019}. 
Circular dipoles provide perhaps the simplest example of 
the intrinsic coupling of spin and orbital angular momentum (SAM and OAM, respectively) in non-paraxial light fields
\cite{allen_orbital_1992,bliokh_transverse_2015,enk_commutation_1994,dorn_focus_2003,nieminen_angular_2008,monteiro_angular_2009,bliokh_angular_2010,rodriguez-herrera_optical_2010,bliokh_spin--orbital_2011,angelsky_orbital_2012}. 
Such fields, in the form of tightly focused laser beams, are of central importance in a rapidly growing range of experiments involving (arrays of) optical tweezers 
\cite{thompson_coherence_2013,barredo_atom-by-atom_2016,endres_atom-by-atom_2016,bernien_probing_2017,cooper_alkaline-earth_2018,norcia_microscopic_2018,pagano_fast_2019,jackson_number-resolved_2020,anderegg_optical_2019,saskin_narrow-line_2019,kun-peng_wang_high-fidelity_2020}. 
These are developed as precise tools to hold and manipulate single atoms or molecules at the quantum level, in creating platforms for quantum simulation and computation, as well as for quantum sensing and atomic clocks \cite{madjarov_atomic-array_2019,norcia_seconds-scale_2019}.

\begin{figure}[!b]
\centering
\includegraphics[width=0.9\columnwidth]{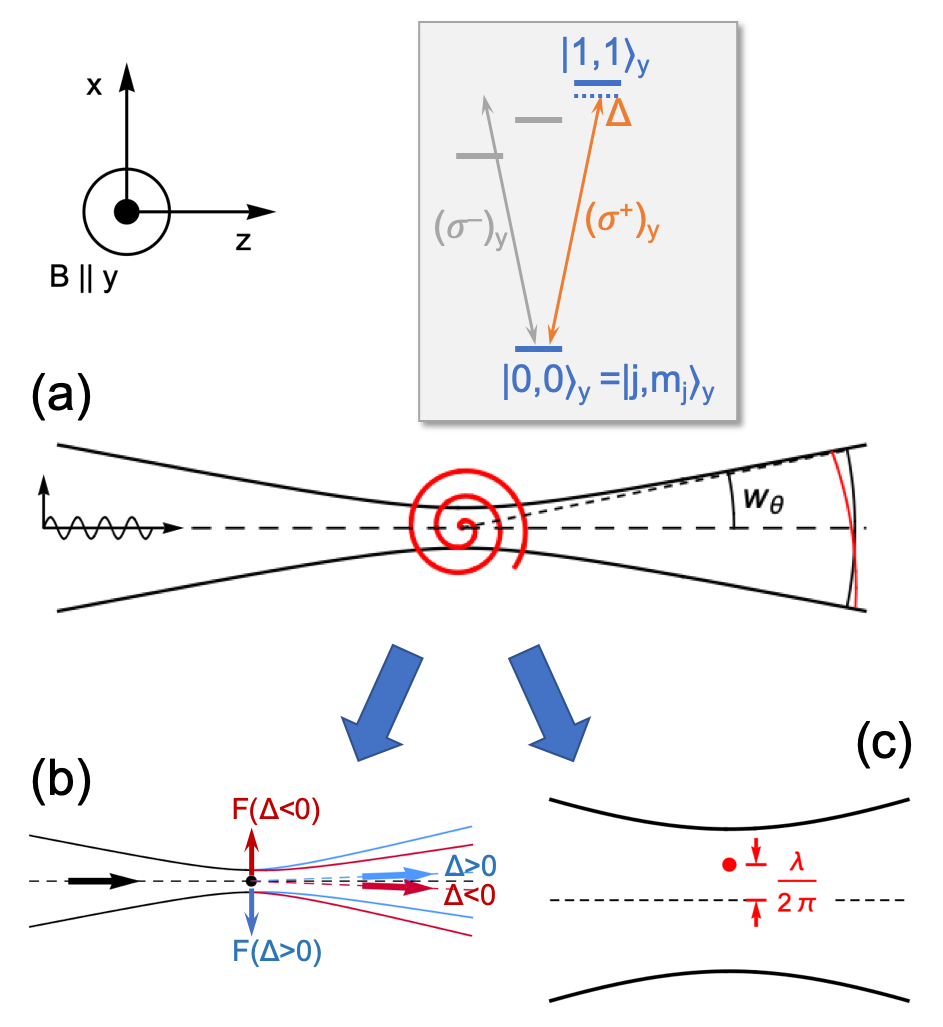}
\caption[]{Optical analog of the Magnus effect. (a) A linearly polarized ($\mathbf{E}\parallel x$), focused laser induces a circular dipole ($xz$ plane) on a $j=0\rightarrow j'=1$ ($\Delta m_j=1$) transition, with a magnetic field $\mathbf{B}\parallel y$ setting the quantization axis. The spiral wave scattered by the circular dipole interferes with the  incident wave, producing two effects: (b) The incident beam is  deflected in the $xz$ plane, with corresponding reaction force on the atom, transverse to the beam. The direction changes sign with the detuning from the atomic resonance. (c) In an optical tweezer (`red' detuning, $\Delta<0$), the transverse force shifts the trapping position away from the optical axis by an amount $\lambdabar=\lambda/2\pi$. See Fig.\ \ref{fig:pdrive} for a far off-resonance case.  }
\label{fig:spiral-scattering}
\end{figure}

Here we predict that an atomic circular dipole can deflect the centered focused laser beam that induces it. Conversely,  the atom will experience a transverse force when on-axis
\footnote{See also \cite{dooghin_optical_1992} for a very different kind of analogy of the Magnus effect.}. 
An important consequence of this force can be seen in the off-axis displacement of the trapping potential created by an optical tweezer \cite{kun-peng_wang_high-fidelity_2020}. Thus, rather than `seeing an atom where it is not' \cite{araneda_wavelength-scale_2019,schwartz_conservation_2006,arnoldus_subwavelength_2008}, here we describe its counterpart of  `trapping an atom where the focus is not' 
\cite{thompson_coherence_2013,kun-peng_wang_high-fidelity_2020,caldwell_sideband_2020}.   
A simple  geometric argument based on light scattering  shows that the {\em true} displacement of the trapping potential is in fact a direct consequence of the {\em apparent} displacement of an emitting circular dipole \cite{darwin_notes_1932,araneda_wavelength-scale_2019}. 
Tweezer trap displacements have previously been calculated numerically, for specific beam shapes, in terms of vector and tensor light shifts \cite{thompson_coherence_2013,caldwell_sideband_2020}. The geometric argument given here directly shows that for a circular dipole the displacement is simply $\pm\lambdabar=\pm\lambda/2\pi$, and  is  remarkably independent of many parameters, including the laser detuning, trap frequency, and even the detailed shape of the trap. A comparison of a Gaussian beam  with an angular tophat beam illustrates this. 
This profound insight provides the basis for state-dependent manipulation of  atomic motion within the tweezer.

We describe these effects in terms of interference between the  focused incident beam with the  wave scattered by the circular dipole, see Fig.~\ref{fig:spiral-scattering}. 
In the optical theorem, such interference is used to describe the attenuation of light in terms of the forward scattering amplitude \cite{jackson_classical_1999}. In contrast, here we concentrate on beam deflection, as a  consequence of the tilt of the spiral wavefront with respect to the incident wave.  
Two simple atomic level schemes serve as examples; (i) a  $j=0\rightarrow j'=1$ and (ii) a $j=1\rightarrow j'=0$ transition. Case (i) is conceptually simpler and most suited to observe the Magnus-like deflection of a weak, near-resonant probe beam. Case (ii) offers interesting extra opportunities in the usual far off-resonance regime of optical tweezer experiments.

Starting with case (i), the $j=0\rightarrow j'=1$ transition, we focus a linearly polarized ($\mathbf{E}\parallel x$), monochromatic laser onto a single atom placed in the origin, see Fig.~\ref{fig:spiral-scattering}. A magnetic field $\mathbf{B}\parallel y$ defines the quantization axis and splits the excited state into three $|j,m_j\rangle_y$ sublevels, separated by the Zeeman shift $\sim\mu_B B/\hbar$, with $\mu_B$ the Bohr magneton \cite{schlesser_light-beam_1992}. We tune the laser close to the $\Delta m_j=+1$ transition, with a detuning $\Delta=\omega_L-\omega_0$  small compared to the Zeeman shift, so that the $\Delta m_j=0,-1$ transitions can be neglected (for example, $\Delta/2\pi\sim10\,$MHz and $\mu_B B/h\sim100\,$MHz.) The emission by the induced circular dipole has a spiral wavefront in the $xz$ plane,  tilted with respect to the forward  $\mathbf{\hat{z}}$ direction of the incident beam.

We represent the  light fields by their angular spectrum \cite{mandel_optical_1995,supmat}, using spherical $k$-space  coordinates $(k,\theta,\phi)$. For monochromatic light, with $k=\omega/c$ fixed, the incident field can be written as  $\frac{1}{2}\mathbf{E}_\text{in}(\Omega) e^{-i\omega t}+c.c.$, with   $\Omega=(\theta,\phi)$. The total field is the sum of the incident and scattered waves. Writing only the positive frequency ($\sim e^{-i\omega t}$) components, the total field reads
\beq
	\mathbf{E}(\Omega)=\mathbf{E}_\text{in}(\Omega)+\mathbf{E}_{\rm{sc}}(\Omega)
\label{eq:Etot}
\eeq
with $\mathbf{E}_{\rm{sc}}(\Omega)$ the scattered wave.

We define the radiant intensity, 
\beq
	J(\Omega)=|\mathbf{E}(\Omega)|^2/2Z_0.
\label{eq:JOmega}
\eeq 
with $Z_0=1/\epsilon_0 c$, so that $J(\Omega)d\Omega$ is the power flowing out of an infinitesimal solid angle $d\Omega=\sin\theta\,d\theta\,d\phi$ around 
\beq
	\mathbf{u}_\Omega=(\sin\theta\,\cos\phi,\sin\theta\,\sin\phi, \cos\theta)
\eeq

Combining Eqs.~(\ref{eq:Etot}) and (\ref{eq:JOmega}), the total radiant intensity is the sum of  three terms,
\beq
	J(\Omega)  = J_\text{in}(\Omega)+J_\text{sc}(\Omega) +J_\text{if}(\Omega)
	\label{eq:Jcomponents}
\eeq
The interference term
\beq
	J_\text{if}(\Omega)=\frac{1}{2Z_0}\left[\mathbf{E}^\ast_\text{in}(\Omega)\cdot\mathbf{E}_\text{sc}^\text{(coh)}(\Omega)+c.c.\right]
	\label{eq:JifOmega}
\eeq
contains only the coherent component of the scattered field. An incoherent component would contribute to $J_\text{sc}(\Omega)$ but not to $J_\text{if}(\Omega)$. For simplicity we  assume that the scattered field is entirely coherent, essentially restricting ourselves  to the low-saturation limit \cite{supmat}.

The deflection of the light beam can be expressed as the change in average wave vector $\langle \mathbf{k}\rangle-\langle \mathbf{k}\rangle_\text{in}$ between the  total (incident plus scattered) and the incident wave, using
\beq
	\left\langle \mathbf{k}\right\rangle_\text{in}=k\,\frac{\int \mathbf{u}_\Omega\,J_\text{in}(\Omega)\,d\Omega}{\int J_\text{in}(\Omega)\,d\Omega}=k\,\frac{\int \mathbf{u}_\Omega\,J_\text{in}(\Omega)\,d\Omega}{P_\text{in}}
\eeq
and similar for  $\langle \mathbf{k}\rangle$, omitting the subscript. 
Assuming (again for simplicity) that  non-radiative decay is absent, we shall write $P_\text{in}=P$ throughout.

The deflection is entirely determined by the 
interference term $J_\text{if}(\Omega)$. The scattered light itself does not contribute, due to the symmetry of the dipole radiation pattern, $J_\text{sc}(\theta,\phi)=J_\text{sc}(\pi-\theta,\pi+\phi)$, so that $\int \mathbf{u}_\Omega J_\text{sc}(\Omega)\,d\Omega=0$. For the deflection we therefore have
\beq
	\delta\langle \mathbf{k}\rangle = 
	\langle \mathbf{k}\rangle-\langle \mathbf{k}\rangle_\text{in} =
	\frac{k}{P}\,\int \mathbf{u}_\Omega\,J_\text{if}(\Omega)\,d\Omega
	\label{eq:deltak}
\eeq
and for the force on the atom, by momentum conservation,
\beq
	\mathbf{F}=-\frac{P}{\omega}\,\delta\langle \mathbf{k}\rangle=-\frac{1}{c}\,\int \mathbf{u}_\Omega\,J_\text{if}(\Omega)\,d\Omega
\eeq
While this expression includes  the forward radiation pressure force, in the cases of interest here the main force will be transverse to the optical axis, $\mathbf{F}\approx F_x \mathbf{\hat{x}}$. Then (approximately) $\delta\langle \mathbf{k}\rangle\perp \langle \mathbf{k}\rangle_\text{in}$ and with 
 $\langle \mathbf{k}\rangle_\text{in}\approx k\mathbf{u}_z$  the deflection angle is 
\beq
	|\delta\theta|\approx
	\frac{|\delta\langle \mathbf{k}\rangle|}{k}
	\label{eq:deltatheta}
\eeq
We will choose $\delta\theta>0$ if $F_x<0$.


Let us now introduce specific field patterns to calculate $J_\text{if}(\Omega)$.
We take the  dipole to be circular, $\mathbf{p}=p e^{i\alpha} \mathbf{u}_+$, with $\mathbf{u}_\pm=(\mathbf{\hat{x}}\mp i\mathbf{\hat{z}})/\sqrt{2}$
denoting spherical unit vectors, and $\alpha$ the  phase of the $p_x$ component of the dipole, relative to the local driving field. 
The field radiated by a coherent dipole \cite{jackson_classical_1999}, in angular coordinates, takes the form \cite{supmat}: 
\beq
	\mathbf{E}_\text{sc}(\Omega) =
	\mathcal{E}_\text{sc}\,i e^{i \alpha } \left(\mathbf{u}_\Omega\times\mathbf{u}_+\right) \times\mathbf{u}_\Omega 
\label{eq:EscOmega}
\eeq
with corresponding $J_\text{sc}(\Omega)$ given by Eq.~\eqref{eq:JOmega}. Here  $\mathcal{E}_\text{sc}=p\,k^2/4\pi\epsilon_0>0$ is a real-valued amplitude. Assuming the steady state of the optical Bloch equations for a two-level system, $\cot\alpha=-\Delta/\gamma$,
with $\Delta=\omega-\omega_0$ the detuning from the $\Delta m_j=+ 1$ transition, and $\gamma=\omega_0^3 D^2/6\pi\epsilon_0\hbar c^3$ the half width of the transition, with  $D$ the transition dipole moment.


For comparison, we consider two different  types of incident beams, Gaussian (`G') and `angular tophat' (`$\Pi$'), where the latter approximates the output of a uniformly illuminated focusing lens. The field for these two beams can be written as 
\begin{eqnarray}
	\mathbf{E}^\text{(G)}_\text{in}(\Omega) & \approx & \mathcal{E}_0^\text{(G)}\,\exp[-\theta^2/w_\theta^2]\,\mathbf{u}_x(\Omega) \label{eq:EinG}\\
	\mathbf{E}^{(\Pi)}_\text{in}(\Omega) & = &
 	\mathcal{E}_0^{(\Pi)}\,\Pi(\theta/2r_\theta)\,\mathbf{u}_x(\Omega) \label{eq:EinPi}
\label{eq:EinOmega}
\end{eqnarray}
with amplitudes $\mathcal{E}_0^\text{(G)},\mathcal{E}_0^{(\Pi)}>0$. 
The Gaussian beam has an angular width $w_\theta$   which is related to the minimum waist  $w_0$ ($1/e^2$ spatial radius of intensity) as $w_\theta w_0=\lambda/\pi$. For the angular tophat, $\Pi(\theta/2r_\theta)$ is the rectangular function with angular half width $r_\theta$ and unit amplitude. Its spatial profile near the focus is the familiar Airy disk pattern. 
Note that neither propagation phases nor the Gouy phase are visible here, as the above  expressions are in  angular coordinates \cite{supmat}.

The polarization vector $\mathbf{u}_x(\Omega)$ is transverse to $\mathbf{u}_\Omega$; it is obtained by  co-rotating  $\mathbf{\hat{x}}$ when rotating  $\mathbf{\hat{z}}\rightarrow\mathbf{u}_\Omega$, i.e.\ rotating by $\theta$ around an axis $\mathbf{\hat{z}}\times\mathbf{u}_\Omega$ \cite{richards_electromagnetic_1959,rodriguez-herrera_optical_2010}, 
\beq
	\mathbf{u}_x(\Omega)=
	\begin{pmatrix}
	\cos\theta  \cos^2 \phi +\sin ^2 \phi  \\
	(\cos \theta -1)\sin \phi  \cos \phi  \\
   	-\sin \theta  \cos \phi 
	\end{pmatrix}
\eeq



When combining Eq.~\eqref{eq:EscOmega} with Eq.~\eqref{eq:EinG} or \eqref{eq:EinPi}  in Eq.~\eqref{eq:JifOmega}, the interference term contains the  amplitude product $\mathcal{E}_0^\text{(G)}\mathcal{E}_\text{sc}$
or $\mathcal{E}_0^{(\Pi)}\mathcal{E}_\text{sc}$. In the low-saturation limit, the amplitude $\mathcal{E}_\text{sc}$ is proportional to $ \mathcal{E}_0^\text{(G)}$ or $ \mathcal{E}_0^{(\Pi)}$. Their ratio can be obtained by requiring energy conservation \cite{supmat}. Upon insertion of the resulting ratios $\mathcal{E}_\text{sc}/ \mathcal{E}_0^\text{(G)}$ and $\mathcal{E}_\text{sc}/ \mathcal{E}_0^{(\Pi)}$ into Eq.~\eqref{eq:JifOmega}, the interference term $J_\text{if}(\Omega)$ becomes proportional to the total power;  the deflection angle is then independent of power.

\begin{figure}[t]
\includegraphics[width=\columnwidth]{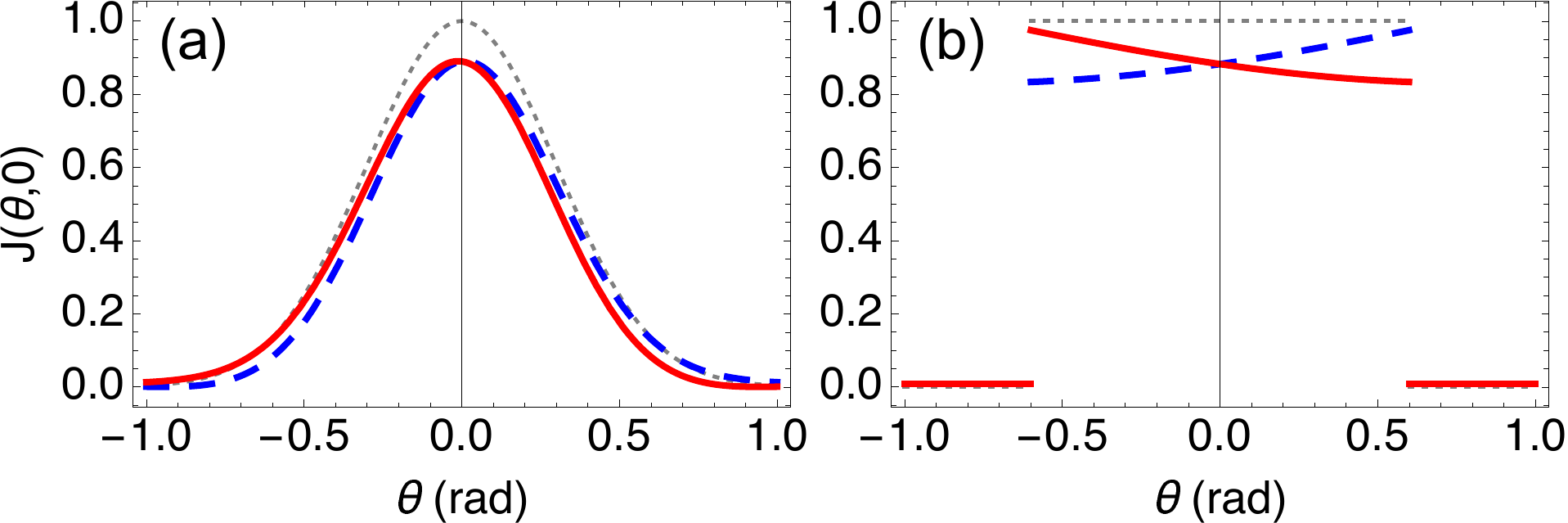}
\caption[]{Beam deflection: radiant intensities in the plane of the $\mathbf{u}_+$ dipole,  for (a) a Gaussian incident beam with $w_\theta=0.6$, and (b) an angular tophat incident beam with $r_\theta=0.6$.
In both cases, the gray/dotted curve shows $J_\text{in}(\theta,\phi=0)$ of the incident beam, normalized to 1 for $\theta=0$; red/solid and blue/dashed curves show the outgoing, or total  $J(\theta,0)$, for $\Delta=-\gamma$ and $+\gamma$, respectively. For clarity, we identify $(\theta,0)\equiv(-\theta,\pi)$. Curves remain the same upon switching simultaneously the signs of the detuning and the spin of the dipole. }
\label{fig:gaussianshift}
\end{figure}

In Fig.~\ref{fig:gaussianshift} we show $J_\text{in}(\Omega)$ in the plane of the dipole ($\phi=0$), together with the total radiant intensity $J(\Omega)$.
For the Gaussian beam, the effect of $J_\text{if}(\Omega)$ is to shift the peak and the average of the direction of propagation away from $\theta=0$. For the angular tophat, the interference leads to an intensity gradient across  the angular  width of the beam, whereas the edges stay at the same angle. In this case the intensity gradient leads to a change in average beam direction.

Finally, the deflection angle is obtained by integration as in Eq.~\eqref{eq:deltak},
{\renewcommand{\arraystretch}{1.5}
\beq
	\delta\theta\approx
	\frac{3}{4}
	\frac{ \gamma  \Delta}{\left(\gamma ^2+ \Delta^2\right)}\times
	\left\{\begin{array}{ll}
   		w_\theta^4 & \quad\text{(Gauss)}\\
		r_\theta^4/4 & \quad\text{(angular tophat)}
	\end{array}\right.
	\label{eq:deflangle}
\eeq} 
and the reaction force as
\beq
	F_x\approx -\frac{P}{c}\delta\theta
\eeq
The results are given as the leading order in $w_\theta$ and $r_\theta$.
The deflection angle  reaches maximal values of $\delta\theta=\pm 3w_\theta^4/8$ and  $\pm 3r_\theta^4/32$, respectively, for $\Delta=\pm\gamma$; it vanishes in the plane-wave limit, $w_\theta, r_\theta\rightarrow 0$.
In this central result we recognize in the  detuning dependence that the force is essentially a dipole force \cite{gordon_motion_1980}, arising from polarization gradients  near the focus of a linearly polarized light beam 
\cite{bliokh_spin--orbital_2011,dorn_focus_2003,
monteiro_angular_2009,nieminen_angular_2008,thompson_coherence_2013,kun-peng_wang_high-fidelity_2020,caldwell_sideband_2020}.

We now address the question of how we can observe the deflection of a laser beam,
either directly or via the reaction force on the atom. 
As shown by  Eq.~\eqref{eq:deflangle},  the angle of deflection by a single atom is small compared to the divergence angle, $|\delta\theta|\ll r_\theta, w_\theta$.
A direct observation  will thus require sufficiently high signal-to-noise ratio, similar to what was achieved in the recent observation of apparent $\lambdabar$  displacement of an emitter \cite{araneda_wavelength-scale_2019}. 
With maximal signal occurring near resonance ($\Delta=\pm\gamma$), where the photon scattering rate is high, the best approach would be to hold the atom in an independent  trap, such as an ion trap or a tight optical tweezer. 
One can then look for the deflection of a weak, near-resonant probe beam. 
A larger deflection angle may be obtained if multiple atoms cooperate. For example, one may consider dense clouds  of sub-wavelength size, containing tens to hundreds of atoms, that have been observed to show collective scattering properties \cite{pellegrino_observation_2014,machluf_collective_2019}. 
Another possibility may be to use elongated, (quasi-) one-dimensional samples with tight ($\lesssim\lambdabar$) radial confinement, achievable, e.g.,  in optical lattices \cite{moritz_exciting_2003,paredes_tonksgirardeau_2004,kinoshita_observation_2004} and on atom chips \cite{jacqmin_sub-poissonian_2011}.

\begin{figure}[t]
\includegraphics[width=\columnwidth]{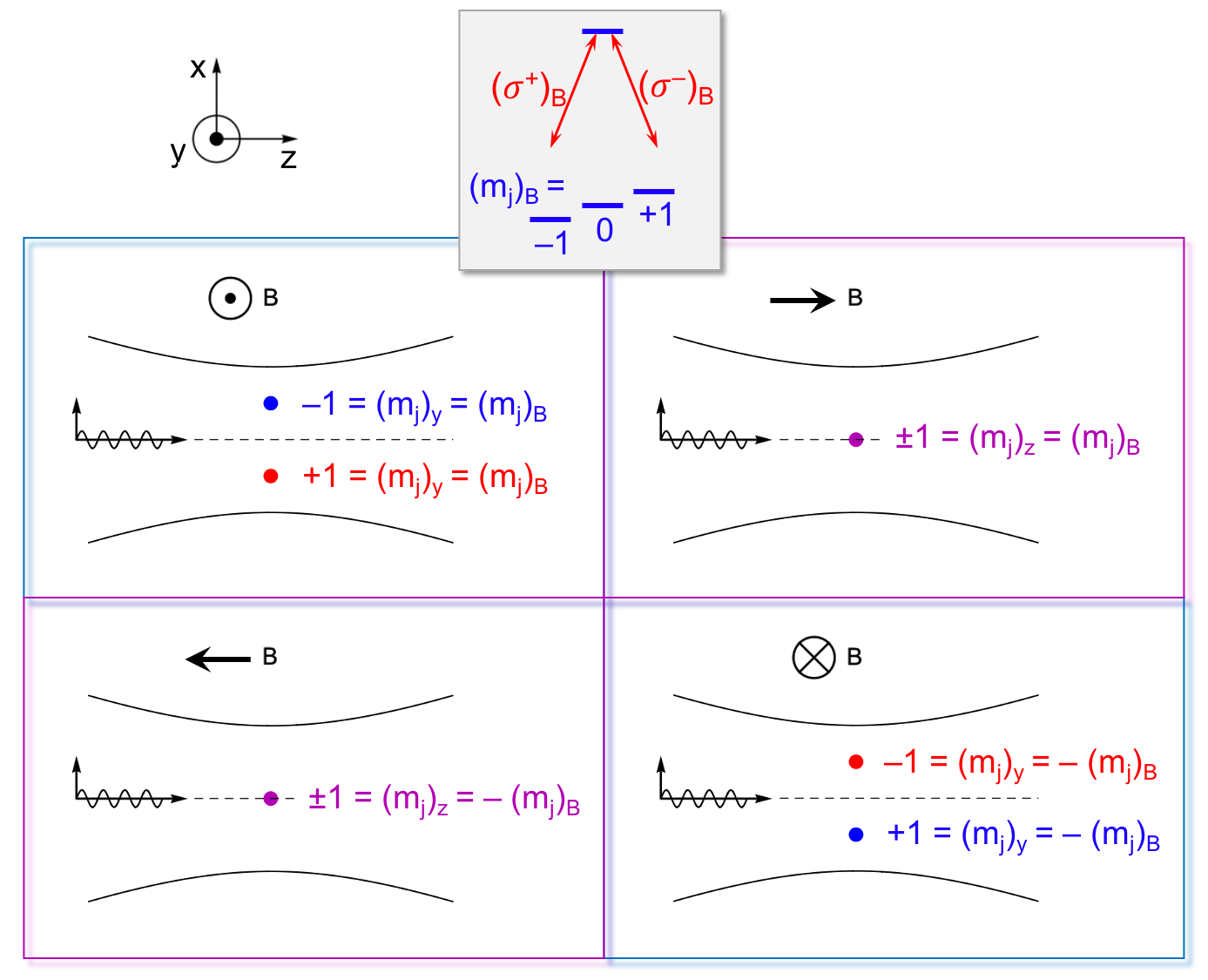}
\caption[]{Optical tweezer operating on a $j=1\rightarrow j'=0$ transition, leading to $\pm\lambdabar$ off-axis displacements for the $(m_j)_y=\mp1$ sublevels  (upper left). The four panels show, in clockwise order, the effect of a rotation of the quantization axis ($\mathbf{B}$), through a cycle $y\rightarrow z\rightarrow -y\rightarrow -z$. While the $\mathbf{B}$-referenced $(m_j)_B$  of an atom is conserved, the space-referenced $(m_j)_y$ is not. The locations of the $(m_j)_B=\pm1$ traps move up and down along the $x$ axis, in antiphase. 
If  $\mathbf{B}$ is rotated at the trap frequency, spin-dependent oscillatory motion in the tweezer can be induced. }
\label{fig:pdrive}
\end{figure}

The second mode of observation, via the force on the atom, provides extra opportunities to manipulate spin-dependent atomic motion in an optical tweezer.
To see this we consider case (ii):  an optical tweezer trapping an atom with a $j=1\rightarrow j'=0$ transition. 
The  $|m_j=\pm1\rangle_y$ states now couple to the $(\sigma^\mp)_y$ components of the light field, and therefore experience opposite forces $F_x$. In this configuration there is no need for a separate probe beam \cite{kun-peng_wang_high-fidelity_2020}, the far off-resonance light ($\Delta/2\pi\sim1-10\,$THz)  of the tweezer itself is sufficient. The  photon scattering and associated heating rates can thus be kept as low as in typical tweezer experiments. 
In this case we assume that the Zeeman shift is large compared to the trap depth $U_0$ (for example $\mu_B B/h\sim10\,$MHz, and $U_0/h\sim1\,$MHz.)
Looking at the spiral wave of a $\mathbf{u}_+$ dipole shown in Fig.~\ref{fig:spiral-scattering}, we can readily see that the relative tilt of the forward wavefronts will vanish if we displace the atom by $\lambdabar$ in the $x$ direction. By thus aligning the wave fronts, the transverse force should vanish. An atom in the 
$|m_j=-1\rangle_y$ sublevel will therefore find an equilibrium position in the tweezer at a displaced off-axis location $x_\text{eq}=\lambdabar$. By the same reasoning, the $|m_j=+1\rangle_y$ sublevel will have the opposite displacement, so that for the $j=1\rightarrow j'=0$ transition:
\beq
	x_\text{eq}=-(m_j)_y \lambdabar
\eeq
The tweezer thus traps the atom off axis, `where the focus is not', in a spin-dependent location. For the situation considered here the $|m_j=0\rangle_y$ state would be untrapped, for a lack of $\pi$ component in the laser polarization. This could be changed by rotating $\mathbf{B}$. In particular,  setting the angle between $\mathbf{E}_\text{in}$ and $\mathbf{B}$ to  $\arctan(\sqrt{2})$, the polarization components  $\sigma^-$, $\pi$, and $\sigma^+$ would become equal. At this `magic angle' all three spin components would be trapped with a   Stern-Gerlach type separation \cite{kun-peng_wang_high-fidelity_2020,stellmer_detection_2011}.

These simple geometric arguments are backed up by a calculation \cite{supmat}, that shows that Eq.~\eqref{eq:deflangle} for the beam deflection is multiplied by $1\mp kd$, for a $\mathbf{u}_\pm$ dipole displaced by $d$ in the $x$ direction, to lowest order in $d$. Thus the transverse force indeed vanishes for a transverse displacement of $d=k^{-1}=\lambdabar$ in the $x$ direction. Remarkably, the size of the displacement is independent of the detuning, the beam divergence angle, the trap frequency, or even the precise shape of the beam (Gauss vs.\ angular tophat). This profound insight follows from the geometric properties of the  scattering problem.

The off-axis trapping  locations offer  interesting opportunities to manipulate the motion of atoms in the tweezer, see Fig.~\ref{fig:pdrive}. Let us imagine an atom trapped in the $|m_j=1\rangle_y$ state. As we  slowly rotate the magnetic field in the $yz$ plane,  the orientation of the atom will adiabatically follow the rotating quantization axis. After rotating the field $y\rightarrow z\rightarrow -y$, the spin will have maintained its orientation relative to $\mathbf{B}$, i.e.\ $|m_j=1\rangle_B\rightarrow|m_j=1\rangle_B$. However, its orientation will have flipped in space,  $|m_j=1\rangle_y\rightarrow|m_j=-1\rangle_y$, since  $\mathbf{B}$ has changed direction. The space-referenced spin flip implies that the atom must have moved to the other side of the optical axis. Thus, by rotating the magnetic field in the $yz$ plane at a frequency $\omega_B$, we effectively shake the trap back and forth: $x_\text{eq}=-(m_j)_B\lambdabar\cos\omega_B t$. The $m_j=\pm 1$ levels are shaken with opposite phase. 

Shaking the trap at an amplitude $\lambdabar$ is equivalent to a harmonic driving force $F_x=m\omega^2\lambdabar\cos\omega_B t$, with $\omega$ the trap frequency. 
Resonant shaking, $\omega\approx\omega_B$,  will induce an oscillatory motion in the trap.  For example, for a tweezer with a laser wavelength of $\lambda\approx0.8\,\mu$m, a Gaussian waist of 2$\,\mu$m, holding an atom of mass $m=88u$ in a  20$\,\mu$K deep trap, the trap frequency will be  $\omega\approx2\pi\times7\,$kHz.  In a simple driven harmonic oscillator model  only 3.5 drive cycles would impart enough energy to kick the atom out of the trap, corresponding to a velocity of $\sim 6\,$cm/s. In reality one would of course need to take anharmonicity into account. The  point here is that  magnetic field modulation can easily induce oscillatory motion in the trap which can then be detected either as trap loss, or by using time-of-flight imaging methods. For the required magnetic field a few gauss should be sufficient, to ensure that the Larmor frequency is large compared to the trap frequency. Rotating the field at frequencies of $\sim10\,$kHz is well possible, being comparable to what is used in TOP traps \cite{petrich_stable_1995}.

Many available atomic level systems should be suitable to display off-axis tweezer trapping. For example, in $^{88}$Sr the transition 
${}^3P_2\rightarrow {}^3S_1$ would provide a $j=2\rightarrow j'=1$ transition. The outer $(m_j)_y=2\;(-2)$ state couples only to the $\sigma^-\;(\sigma^+)$ polarization component, so its spatial shift will be $-\lambdabar\;(+\lambdabar)$.  Using $^{87}$Rb one could 
operate a tweezer red detuned to the  $D_1$ line (795~nm), driving the two hyperfine lines  $F=2\rightarrow F'=1,2$. Also in this case  the outer state $(m_F)_y=2\;(-2)$  is displaced by $-\lambdabar\;(+\lambdabar)$, as long as the detuning stays small compared to the fine structure splitting of the $D$ lines.


In summary, it is predicted  that a circular dipole can deflect a focused laser beam, similar to a spinning ball deflecting a stream of air in the Magnus effect. 
The reaction force on the atom leads  to spin-dependent, off-axis  displacement of atoms trapped in an optical tweezer. For a pure circular dipole the displacement is $\pm\lambdabar$, independent of many trap parameters. An external magnetic field can be used to induce spin-dependent motion or to perform Stern-Gerlach type analysis of the spin states of the atom in the tweezer.

I would like to thank  N.J. van Druten, R. Gerritsma, J. Minar, and A. Urech  for  stimulating  and encouraging discussions as well as careful reading of the manuscript. This work was supported by the Netherlands Organization for Scientific Research (NWO).



%

\clearpage

\widetext

\begin{center}
\textbf{\large Supplemental Materials for:\\Off-axis optical tweezers by an optical analog of the {Magnus} effect}
\\
Robert J.C. Spreeuw\\
\em{Van der Waals-Zeeman Institute, Institute of Physics, University of Amsterdam, \\
PO Box 94485, 1090 GL Amsterdam, The Netherlands}
\end{center}

\setcounter{equation}{0}
\setcounter{figure}{0}
\setcounter{table}{0}
\setcounter{page}{1}
\makeatletter
\renewcommand{\theequation}{S\arabic{equation}}
\renewcommand{\thefigure}{S\arabic{figure}}
\renewcommand{\bibnumfmt}[1]{[S#1]}
\renewcommand{\citenumfont}[1]{S#1}


\section{Spatial vs. angular $k$-space coordinates}

In this paper we express  all  fields by their angular spectrum $\mathbf{E}(\Omega)=\mathbf{E}(\theta,\phi)$. This is usually defined for fields propagating out into a half space $z\geq 0$ (see Ch.~3.2 in \cite{sm-mandel_optical_1995}), as is clearly the case for the  incident beams. 
The relationship with the field in the plane $z=0$ is given by 
\beq
	\mathbf{\tilde{E}}(x,y,0)=k^2\iint_{\theta\leq\pi/2}
	\mathbf{E}(\Omega)\sin\theta \,d\theta d\phi
\label{eq:ErEOmega}
\eeq
with $k=\omega/c$ the laser wave vector. 
For the Gaussian beam with angular waist $w_\theta$ the above equation yields the familiar Gaussian beam cross section, with minimum waist $w_0=\lambda/\pi w_\theta$, see also Ch.~5 in \cite{sm-mandel_optical_1995}.  
The angular tophat beam approximates the output of a uniformly illuminated circular lens, and  Eq.~\eqref{eq:ErEOmega} yields the resulting Airy pattern in the focal plane $z=0$.

Although the emission by a dipole is not confined to $z\geq0$, the angular representation of the radiation pattern of a dipole $\mathbf{p}$ in a direction $\mathbf{u}_\Omega$ is well known to be given by Eq.~(10) (main text), see for example Ch.~9 in \cite{sm-jackson_classical_1999}. Only the radiating, or `far field', terms ($\sim r^{-1}$) are relevant in our case, because one can evaluate the beam deflection at arbitrarily large distance of the dipole, where the near fields ($\sim r^{-2}, r^{-3}$) have become negligible.

In the plane ($\phi=0$) of a $\mathbf{u}_\pm$ dipole,
\beq
	\left(\mathbf{u}_\Omega\times\mathbf{u}_\pm\right) \times\mathbf{u}_\Omega=\frac{e^{\pm i\theta}}{\sqrt{2}}\left(\cos\theta,0,-\sin\theta\right)
\eeq
shows the spiral wave character in the prefactor $e^{\pm i\theta}$.

The factor $i$ in  Eq.~(10) (main text)  is a  crucial detail. It is a consequence of expressing the spherical waves $e^{ikr}/r$ of the dipole field as an  angular spectrum of plane waves. The same factor $i$ can be recognized in the Weyl representation of a diverging spherical wave \cite{sm-mandel_optical_1995}. 
In the case at hand, one can readily see that it also  ensures that a resonant beam is attenuated (absorbed) in the forward direction, due to destructive interference of incident and scattered waves.

The phase factor $e^{i\alpha}$ in Eq.~(10) (main text)  follows from the steady state of the optical Bloch equations \cite{sm-cohen-tannoudji_atom-photon_1998}. In a two-level atom with states $e,g$, the induced dipole moment is given by the off-diagonal density matrix element $\rho_{eg}$. If the atom is driven at detuning $\Delta$ by a monochromatic  field with (real) amplitude  $\mathcal{E}_0$
the steady state   (for $s\ll 1$) is given by 
\beq
	\rho_{eg}=\frac{i}{2}\frac{D\mathcal{E}_0/\hbar}{\gamma-i\Delta}
\eeq
which has a  complex argument $\alpha=\arg\rho_{eg}$  given by $\cot\alpha=-\Delta/\gamma$.
Here, since we choose an $x$ polarized incident wave, $\alpha$ is the phase of the $p_x$ component of the dipole, relative to the incident field.

\section{Low saturation limit}

In the definition of the saturation parameter $s$ we include the detuning, following \cite{sm-cohen-tannoudji_atom-photon_1998}, 
\beq
	s = \frac{I/I_0}{1+\Delta^2/\gamma^2}
\eeq
with $I$ the intensity and $I_0=2\pi h c\gamma/3\lambda^3$ the saturation intensity. In the  low-saturation limit,  characterized by $s\ll 1$, the scattered light is almost entirely coherent, with a small incoherent fraction equal to $s/(1+s)$.
In optical tweezer experiments, using far off-resonant laser beams, typical values for $s$ are in the range $10^{-6}-10^{-8}$, so that $s\ll 1$ is indeed well fulfilled and the incoherent scattering rate is low.

\section{Field amplitudes}

The peak amplitudes $\mathcal{E}_0^\text{(G)}, \mathcal{E}_0^{(\Pi)}$ are related to the total  power in the incident beam by 
{\renewcommand{\arraystretch}{2}
\beq
	P=\int{J_\text{in}(\Omega)\;d\Omega} =
	\left\{\begin{array}{ll}
   		\approx\frac{(\mathcal{E}_0^\text{(G)})^2}{2Z_0}\times\frac{\pi w_\theta^2}{2}\\
		\frac{(\mathcal{E}_0^{(\Pi)})^2}{2Z_0}\times 2 \pi \left( 1-\cos r_\theta\right)
	\end{array}\right.
\label{eq:Pin}
\eeq}
for the Gaussian and angular tophat beam, respectively. The integrals were performed using Mathematica software \cite{sm-Mathematica}.
For the Gaussian, the equality is only approximate, we give here the leading term of a power series in  $w_\theta$. The above expressions have been written as a product of the forward ($\theta=0$) radiant intensity $\mathcal{E}_0^2/2Z_0$ and an effective solid angle.

The average wave vector of the incident beams is shorter than the corresponding value for a plane wave, 
{\renewcommand{\arraystretch}{1.5}
\beq
	\langle\mathbf{k}\rangle_\text{in}=k\mathbf{\hat{z}}\times 
	\left\{\begin{array}{ll}
   		1-\frac{w_\theta^2}{4}+\mathcal{O}(w_\theta^4) & \text{(Gauss)}\\
		\cos^2 \left(\frac{r_\theta}{2}\right) & \text{(tophat)}
	\end{array}\right.
\eeq}

The amplitude ratios $\mathcal{E}_\text{sc}/ \mathcal{E}_0^\text{(G)}$ and $\mathcal{E}_\text{sc}/ \mathcal{E}_0^{(\Pi)}$ can be obtained from the energy conservation condition
\beq
	\int \left[J_\text{if}(\Omega)+J_\text{sc}(\Omega)\right] \,d\Omega=0
\label{eq:econs}
\eeq
The scattering term $J_\text{sc}(\Omega)>0$ would increase the outflowing power, which must be cancelled by the interference term $J_\text{if}(\Omega)$.
As a result, 
\begin{eqnarray}
	\frac{\mathcal{E}_\text{sc}}{\mathcal{E}_0^\text{(G)}} & \approx & \frac{3   \sin \alpha }{4
   \sqrt{2}}\,w_\theta^2 \\
   	\frac{\mathcal{E}_\text{sc}}{\mathcal{E}_0^{(\Pi)}} & = & 
	\frac{3 \sin \alpha }{4 \sqrt{2}}\,\sin ^2\left(\frac{r_\theta}{2}\right)
   (\cos r_\theta+3)
\end{eqnarray}
where in the Gaussian case the leading order in $w_\theta$ is given. 

With these ratios the interference terms in the radiant intensity, Eq.~(5) (main text) can be obtained as 
{\renewcommand{\arraystretch}{1.5}
\beq
	J_\text{if}(\Omega)=-\frac{\mathcal{E}_\text{sc}  }{\sqrt{2} Z_0 }\,f(\Omega,\Delta)\times
	\left\{\begin{array}{ll}
   		\sim\mathcal{E}_0^\text{(G)} e^{-\theta ^2/w_\theta^2}\\
		\mathcal{E}_0^{(\Pi)} \,\Pi(\theta/2r_\theta)
	\end{array}\right.
\label{eq:Jif}
\eeq}
with 
\beq
	f(\Omega,\Delta) = \frac{\gamma\left(  \cos \theta \cos ^2\phi +  \sin ^2\phi\right) -
   \Delta  \sin \theta  \cos \phi }{
   \sqrt{\gamma ^2+ \Delta ^2}}
\eeq

For the deflection, expressed as $\delta\langle\mathbf{k}\rangle=\langle\mathbf{k}\rangle-\langle\mathbf{k}\rangle_\text{in}$ we evaluate the integral of Eq.~(7) (main text) to obtain
\beq
	\delta\langle\mathbf{k}\rangle=
	\frac{3k}{4}\frac{\gamma\Delta}{\gamma^2+\Delta^2}\,
	\left(w_\theta^4+\mathcal{O}(w_\theta^5),0,-2\frac{\gamma}{\delta}w_\theta^2+\mathcal{O}(w_\theta^4)\right)
\eeq
for the Gaussian beam, and 
\beq
	\delta\langle\mathbf{k}\rangle=
	\frac{3k}{16}\frac{\gamma\Delta}{\gamma^2+\Delta^2}\,
	\left(r_\theta^4+\mathcal{O}(r_\theta^5),0,-4\frac{\gamma}{\delta}r_\theta^2+\mathcal{O}(r_\theta^4)\right)
\eeq
for the angular tophat beam. 

Note that in both cases the $y$ component is absent. To leading order, the deflection angle is just given by 
\beq
	\delta\theta\approx\frac{(\delta\langle\mathbf{k}\rangle)_x}{k}
\eeq
which leads to Eq.~(14) of the main text.

\section{Calculation for a displaced atom}

When the atom is located at a position $\mathbf{d}$ away from the origin, the angular 
components of the scattered wave are phase shifted by an amount $\exp({-ik\mathbf{u}_\Omega\cdot\mathbf{d}})$, so that 
the interference term, Eq.~(5) (main text),
is modified to 
\beq
	J_\text{if}(\Omega)=\frac{1}{2Z_0}\left[\mathbf{E}^\ast_\text{in}(\Omega)\cdot\mathbf{E}_\text{sc}^\text{(coh)}(\Omega)e^{-ik\mathbf{u}_\Omega\cdot\mathbf{d}}+c.c.\right]
	\label{eq:JifOmegaDisp}
\eeq
For a displacement along $x$, we have  $\mathbf{d}=d\mathbf{\hat{x}}$ so that 
\beq
	k\mathbf{u}_\Omega\cdot\mathbf{d}=kd\,\mathbf{u}_\Omega\cdot\mathbf{\hat{x}}=kd\sin\theta\cos\phi 
\eeq
In the 
integrals  $\int J_\text{if}(\Omega)\,d\Omega$ and $\int \mathbf{u}_\Omega\,J_\text{if}(\Omega)\,d\Omega$, we develop the integrand in a power series of $kd$, up to fourth order and integrate the terms separately. 

For the amplitude ratios $\mathcal{E}_\text{sc}/ \mathcal{E}_0^\text{(G)}$ and $\mathcal{E}_\text{sc}/ \mathcal{E}_0^{(\Pi)}$ we find that their lowest order ($\sim w_\theta^2$ and $\sim r_\theta^2$) is not affected by $kd$. 
For the deflection angle the leading order in $w_\theta, r_\theta$ is still fourth order, and  up to  order $(kd)^4$ the angle is 
{\renewcommand{\arraystretch}{1.5}
\beq
	\delta\theta\approx
	\frac{3}{4}
	(1\mp kd)
	\frac{ \gamma  \Delta}{\left(\gamma ^2+ \Delta^2\right)}\times
	\left\{\begin{array}{ll}
   		w_\theta^4 & \quad\text{(Gauss)}\\
		r_\theta^4/4 & \quad\text{(tophat)}
	\end{array}\right.
\eeq}
for a $\mathbf{u}_\pm$ dipole, respectively.
This shows that the deflection angle, and thus also the transverse force, vanishes if the atom is displaced by an amount $d=\pm k^{-1}=\pm\lambdabar$.


%

\end{document}